\documentclass[12pt]{iopart}

\usepackage{cite}
\usepackage{graphicx}
\usepackage{color}

\begin{document}

\title[Resonances for a central--field model]{Accurate calculation of
resonances for a central--field model potential}
\author{Francisco M. Fern\'{a}ndez}

\address{\ddag\ INIFTA (UNLP, CCT La Plata-CONICET), Divisi\'on Qu\'imica Te\'orica,
Blvd. 113 S/N,  Sucursal 4, Casilla de Correo 16,
1900 La Plata, Argentina}\ead{fernande@quimica.unlp.edu.ar}

\maketitle

\begin{abstract}
We obtain accurate resonance energies for the Schr\"{o}dinger equation with
a central--field potential by means of a method based on a rational
approximation to the logarithmic derivative of the wavefunction. We discuss
the rate of convergence of our approach and compare present results with
those obtained earlier by other authors. We show that present method is
superior to the spherical--box approach applied recently to the same
problem. As far as we know present results are more accurate than those
available in the literature and may be a suitable benchmark for testing
future approaches.
\end{abstract}

\section{Introduction}

\label{sec:intro}

In a recent paper Zhou et al\cite{ZMZ09} applied the well--known
spherical--box stabilization method to the calculation of the resonance
energies of the Schr\"{o}dinger equation with the potential $%
V(r)=V_{0}r^{2}e^{-r}+Z/r$. They integrated the eigenvalue equation by means
of the Runge--Kutta method and estimated the positions and widths of the
resonances from the behaviour of the bound--state energies as functions of
the box radius. In particular they calculated the first two s--wave the
first p--wave and the first d--wave resonances. Zhou et al\cite{ZMZ09}
experienced some difficulties in estimating the position an width of the
second s--wave resonance and could not obtain the third one. They also
obtained rather crude estimations of the first p--wave and d--wave
resonances. It seems that the spherical--box approach is rather ill--suited
to broad resonances.

The potential $V(r)$ mentioned above has proved a suitable benchmark for the
development and testing of several methods for the calculation of the
energies of metastable states\cite
{BBJS74,G76,IMM78,J80,MCD80,KLM81,J82,KLM82,MW82,CS83,M84,L89,MRT93,RM93,SE93,F95,LC95,RM95,YA95,CH96,DP97,SR97,TON98}%
. Most authors have considered the case $Z=0$\cite
{BBJS74,CH96,CS83,DP97,F95,G76,IMM78,J80,J82,KLM81,KLM82,L89,LC95,MCD80,MRT93,M84,MW82,RM93,RM95,SE93,SR97,TON98,YA95}
and just a few ones included the Coulomb interaction $Z=-1$\cite{SR97, YA95}.

Some time ago, we applied the Riccati--Pad\'{e} method (RPM) to the
calculation of the lowest s--wave resonance of the potential $V(r)$ with $%
Z=0 $\cite{F95}. In that earlier paper we did not discuss the rate of
convergence of the method on this particular model and merely showed the
result for the lowest s--wave resonance. The purpose of this paper is to
compare the RPM with the spherical--box approach and with other alternative
methods on the central--field potential with $Z=-1$.

In Sec.~\ref{sec:method} we outline the method. In Sec.~\ref{sec:results} we
apply the RPM to the potential $V(r)$ with $Z=-1$, analyze its results and
compare them with those obtained earlier by other authors. Finally, in Sec.~%
\ref{sec:conclusions} we discuss the advantages of the RPM and draw
conclusions.

\section{The method}

\label{sec:method}

The radial part of the dimensionless Schr\"{o}dinger equation for a
central--field potential $V(r)$ is
\begin{equation}
\left[ -\frac{1}{2}\frac{d^{2}}{dr^{2}}+\frac{l(l+1)}{2r^{2}}+V(r)\right]
\Phi (r)=E\Phi (r)  \label{eq:Schrödinger}
\end{equation}
where $l=0,1,\ldots $ is the angular--momentum quantum number and $\Phi (0)=0
$. The RPM is based on a rational approximation to the regularized
logarithmic derivative of the wavefunction
\begin{equation}
f(r)=\frac{l+1}{r}-\frac{\Phi ^{\prime }(r)}{\Phi (r)}  \label{eq:f(r)}
\end{equation}
that can be expanded as follows:
\begin{equation}
f(r)=\sum_{j=0}^{\infty }f_{j}r^{j}  \label{eq:f_series}
\end{equation}
Note that the term $(l+1)/r$ removes the singularity of $\Phi ^{\prime
}(r)/\Phi (r)$ at origin and that we can obtain the coefficients $f_{j}(E)$
analytically by means of simple recurrence relations\cite{F95}.

We then convert the Taylor series into a rational approximation:
\begin{equation}
\left[ N+d/N\right] =\frac{\sum_{j=0}^{N+d}a_{j}r^{j}}{%
\sum_{j=0}^{N}b_{j}r^{j}}=\sum_{j=0}^{2N+d+1}f_{j}r^{j}  \label{eq:Padé}
\end{equation}
where $N=1,2,\ldots $ and $d=0,1,\ldots $. The $2N+d+1$ adjustable
coefficients $a_{j}$ and $b_{j}$ are insufficient to provide the $2N+d+2$
coefficients $f_{j}$. This condition is satisfied only if the Hankel
determinant $H_{D}^{d}(E)$ with matrix elements $f_{i+j+d-1}(E)$, $%
i,j=1,2,\ldots ,D=N+1$, vanishes\cite{F95} (and references therein). The RPM
conjecture is that there are sequences of roots $E^{[D,d]}$ of $%
H_{D}^{d}(E)=0$, $D=2,3,\ldots $ that converge towards the actual bound--
and metastable--state energies of the Schr\"{o}dinger equation (\ref
{eq:Schrödinger}). The calculation is remarkably simple because the Hankel
determinants are polynomial functions of the energy.

\section{Results and discussion}

\label{sec:results}

For comparison we consider the potential
\begin{equation}
V(r)=V_{0}r^{2}e^{-r}+\frac{Z}{r}  \label{eq:V(r)}
\end{equation}
with the model parameters $V_{0}=7.5$ and $Z=-1$\cite{ZMZ09,YA95,SR97}.  We label
the complex energies of the metastable states $E_{l,\nu }$ so
that $\mathop{\rm Re}E_{l,\nu +1}>\mathop{\rm Re}%
E_{l,\nu }$, $\nu =0,1,\ldots $. The s--, p-- and d--waves discussed by Zhou
et al\cite{ZMZ09} and Sofianos and Rakityansky\cite{SR97} correspond to $l=0$%
, $l=1$, and $l=2$, respectively.

Since we are looking for Siegert pseudo states that satisfy\cite{S39}
\begin{equation}
\lim_{r\rightarrow \infty }\frac{\Phi ^{\prime }(r)}{\Phi (r)}=ik
\label{eq:Siegert}
\end{equation}
then it seems reasonable to choose $d=0$ because
\begin{equation}
\lim_{r\rightarrow \infty }[N/N]=\frac{a_{N}}{b_{N}}  \label{eq:lim[N/N]}
\end{equation}
For that reason it should be assumed that $d=0$ from now on, unless
otherwise stated.

\begin{figure}[H]
\begin{center}
\bigskip\bigskip\bigskip \includegraphics[width=9cm]{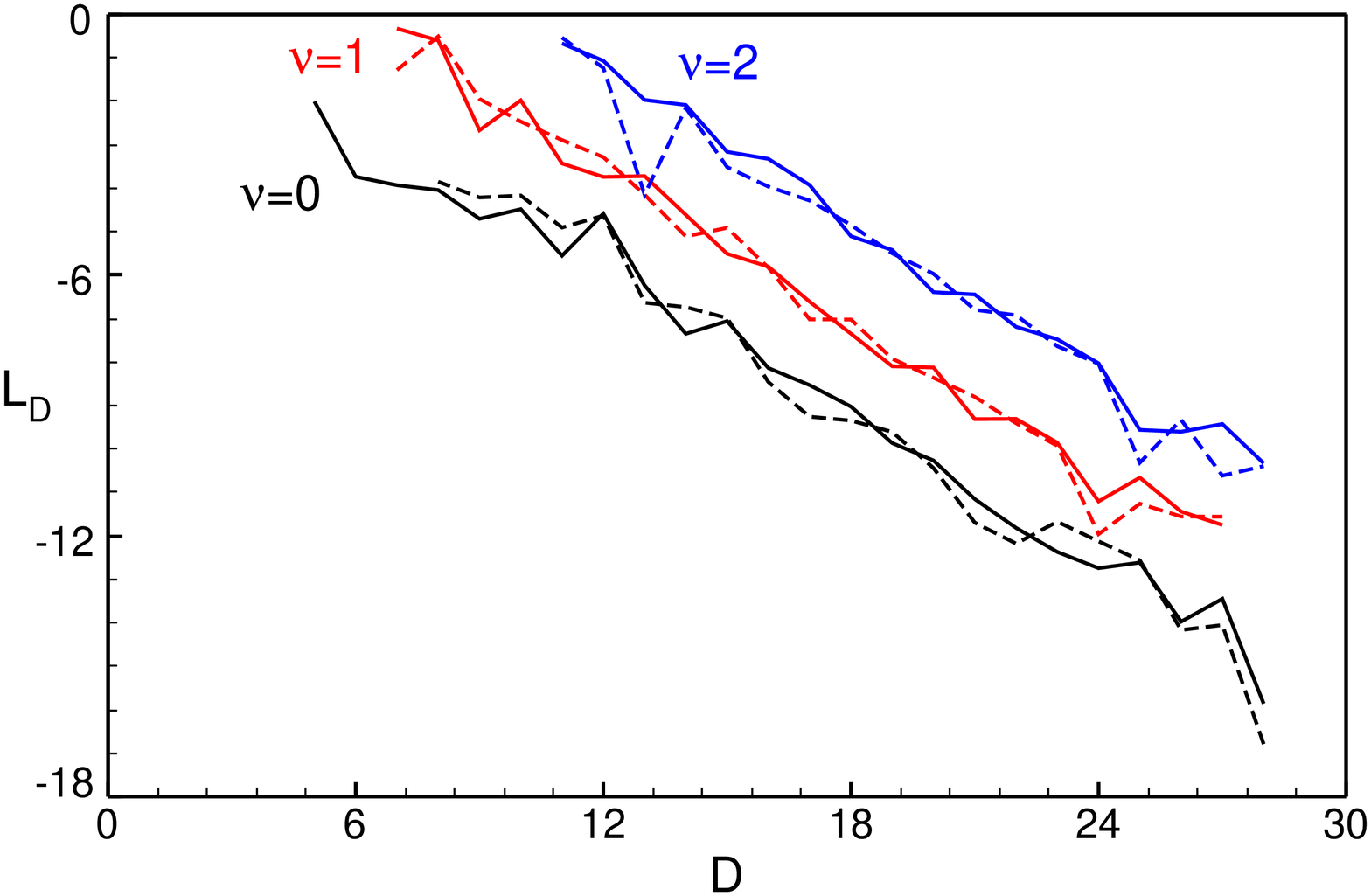}
\end{center}
\caption{Convergence rate $L_D$ for the real (solid line) and imaginary
(dashed line) parts of the energies $E_{0,\nu}$}
\label{Fig:S}
\end{figure}

In order to estimate the rate of convergence of the RPM we calculate $%
L_{D}=\log |\alpha _{D}-\alpha _{D+1}|$ where $\alpha _{D}$ is either the
real or imaginary part of $E^{[D,0]}$. Fig.~\ref{Fig:S} shows that the rate
of convergence of the RPM for both the positions and widths of the first
three s--wave resonances is remarkable. It is worth mentioning that in the
case of a narrow resonance the imaginary part of the root will appear at
sufficiently large determinant dimensions $D$; that is to say, when $|%
\mathop{\rm Re}E^{[D,0]}-\mathop{\rm Re}E^{[D+1,0]}|$ is of the order of
magnitude of $|\mathop{\rm Im}E|$. We appreciate that the rate of
convergence (given approximately by the slope of $L_{D}$) is almost
independent of $\nu $; the main difference is that the greater the value of $%
\nu $ the larger the determinant dimension $D$ necessary for the appearance
of the corresponding sequence. On the other hand, the performance of the
spherical--box approach deteriorates as the resonance width increases\cite
{ZMZ09}. It is clear that the RPM is preferable to the spherical--box
approach, at least for this example.

\begin{figure}[H]
\begin{center}
\bigskip\bigskip\bigskip \includegraphics[width=9cm]{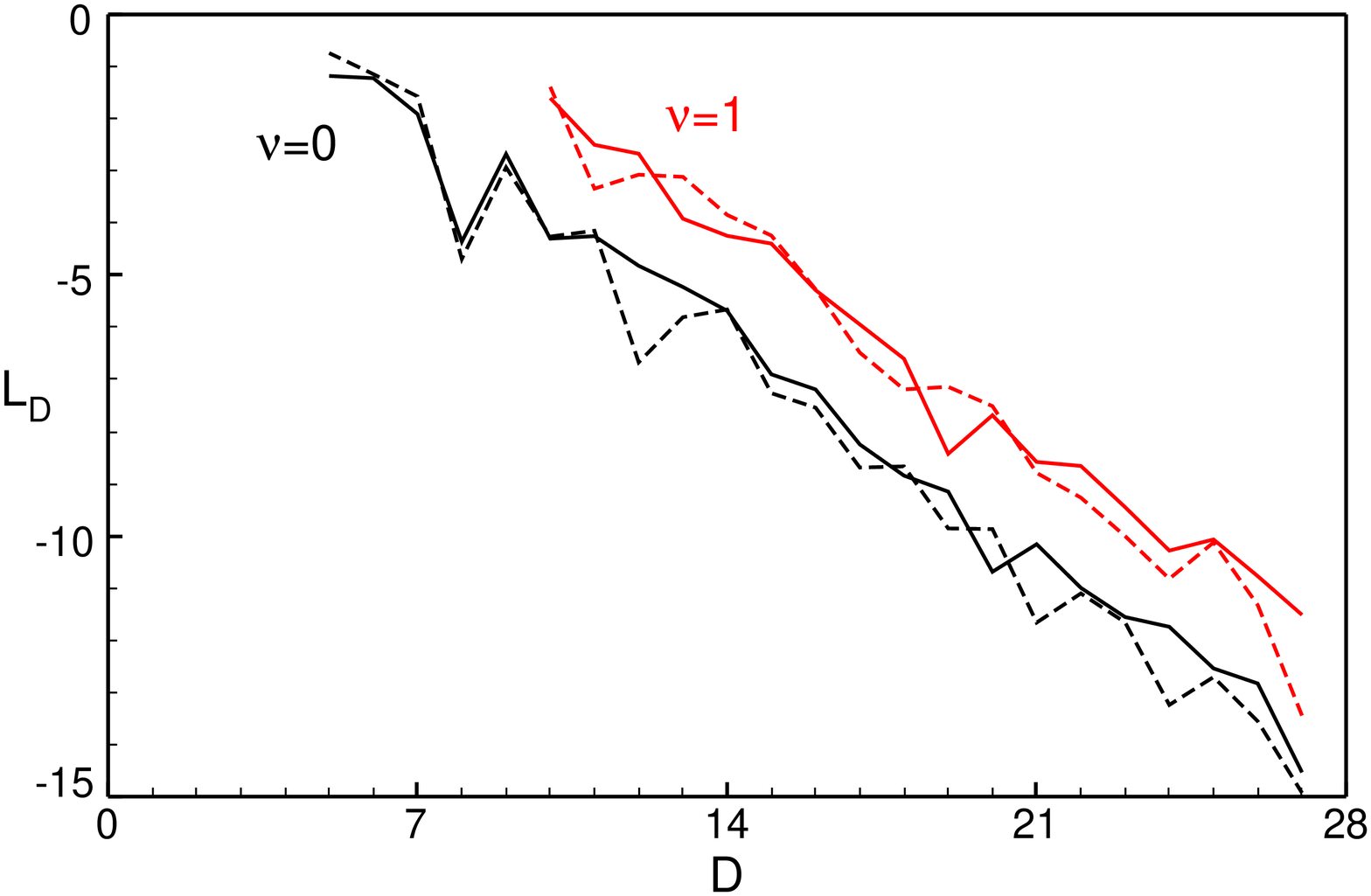}
\end{center}
\caption{Convergence rate $L_D$ for the real (solid line) and imaginary
(dashed line) parts of the energies $E_{1,\nu}$}
\label{Fig:P}
\end{figure}

Fig.~\ref{Fig:P} shows that the rate of convergence of the RPM for the
p--wave resonances ($l=1$) is similar to that discussed above. It is clear
that the rate of convergence of the Hankel sequences is also independent of $%
l$. We confirm our conclusion that the RPM is preferable to the
spherical--box approach because Zhou et al\cite{ZMZ09} roughly estimated the
position and width of $E_{1,0}$ and were unable to obtain other p--wave
resonances.

\begin{figure}[H]
\begin{center}
\bigskip\bigskip\bigskip \includegraphics[width=9cm]{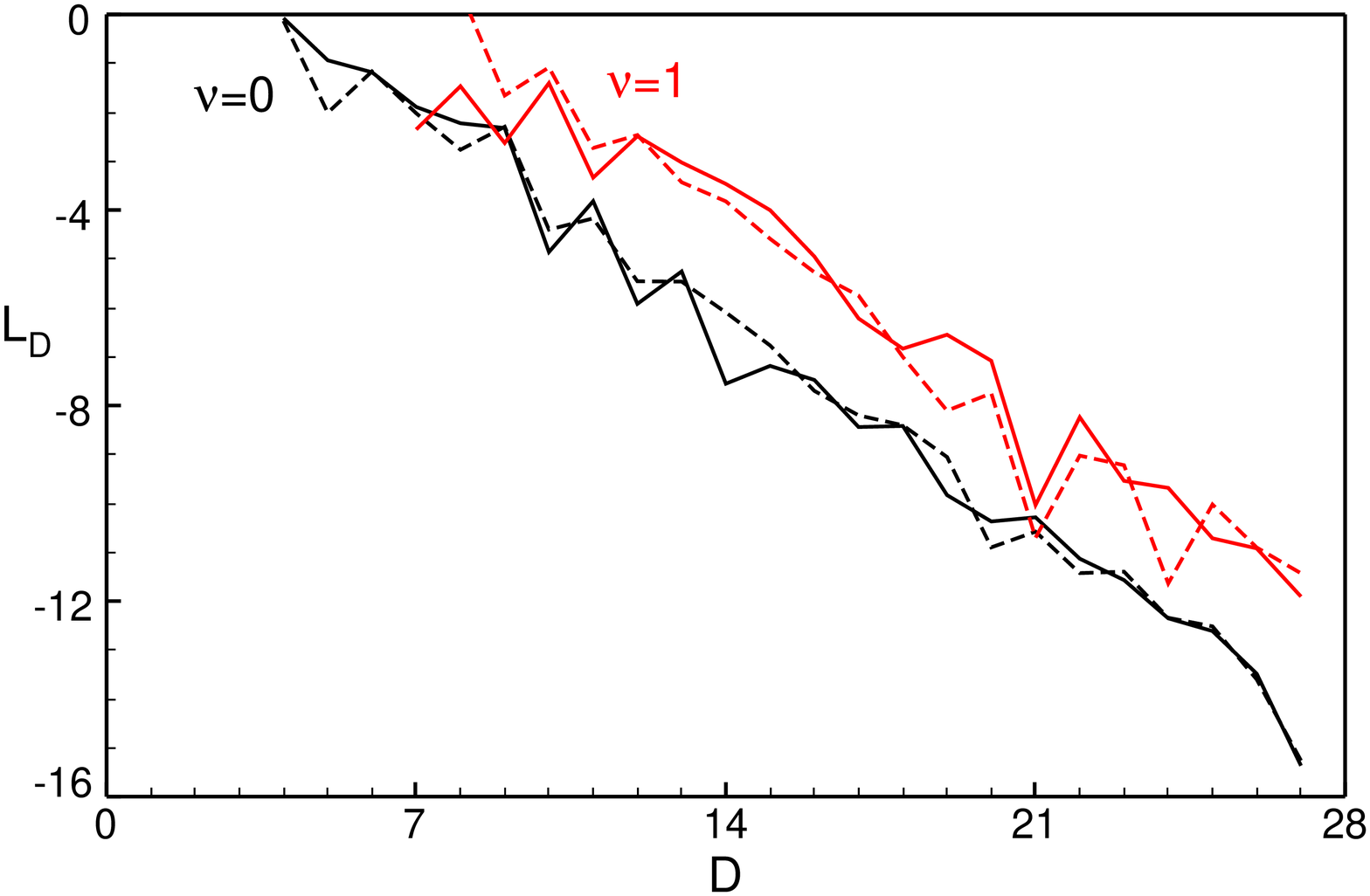}
\end{center}
\caption{Convergence rate $L_D$ for the real (solid line) and imaginary
(dashed line) parts of the energies $E_{2,\nu}$}
\label{Fig:D}
\end{figure}

Fig.~\ref{Fig:D} shows the rate of convergence for the first two d--wave
resonances. The behaviour is similar to those discussed above for the s and
p ones. According to Zhou et al\cite{ZMZ09} the spherical--box approach only
revealed the first d resonance for which they could provide a rather crude
estimate of the position and width.

As far as we know, the most accurate results for this model are those
calculated some time ago by Sofianos and Rakityansky\cite{SR97}. Present
results are even more accurate and may therefore be a useful benchmark for
other approaches. For that purpose we show them in Table~\ref{tab:E}.

Finally, we mention that the rate of convergence of the\ RPM is not affected
by the choice of the displacement $d$. In the present case, for example, we
obtained similar results with $d=1$ that we do not show here.

\section{Conclusions}

\label{sec:conclusions}

One of the main advantages of the RPM is its remarkable simplicity. We first
obtain the coefficients of the Taylor series (\ref{eq:f_series}) by means of
a straightforward recurrence relation\cite{F95}. Second, we construct the
Hankel determinant, which is a polynomial function of the energy, and find
its roots. Third, we identify the sequences of roots that converge towards
physically acceptable results.

Another advantage of the RPM is that exactly the same Hankel determinant
applies to both the bound states and resonances. It comes from the fact that
the RPM does not take explicitly into account the asymptotic form of the
wavefunction at infinity and the rational approximation applies to any
solution of the Schr\"{o}dinger equation. Of course, we have to take into
consideration the behaviour of the wavefunction at origin in order to remove
any singularity of $\Phi ^{\prime }(r)/\Phi (r)$.

We think that present results clearly show that the RPM is much more
accurate and reliable than the spherical--box approximation. We have
calculated the resonances discussed by Zhou et al\cite{ZMZ09} with much more
accuracy and also obtained others that those authors were unable to locate.
Besides, it is worth noting that the RPM is as simple, or even simpler, than
the box--stabilization method in any of its forms\cite{L89, MCD80, MRT93,
ZMZ09}.

\begin{table}[H]
\caption{Some complex energies $E_{l,\nu}$ for the potential~(\ref{eq:V(r)})}
\label{tab:E}
\begin{center}
\begin{tabular}{llll}
\hline
$l$ & $\nu$ & \multicolumn{1}{c}{$\mathop{\rm Re} E$} & \multicolumn{1}{c}{$%
\mathop{\rm Im} E=\Gamma/2$} \\ \hline
0 & 0 & 1.7805245363623048 & 0.00004785969842876 \\
0 & 1 & 4.101494946209 & 0.578627213766 \\
0 & 2 & 4.6634610967 & 2.6832007703 \\
1 & 0 & 3.848001634811759 & 0.137692229585768 \\
1 & 1 & 4.750053489274 & 1.75278992436148 \\
2 & 0 & 4.9005161468291143 & 0.7837535082665858 \\
2 & 1 & 5.3006134902578 & 2.942357430621 \\ \hline
\end{tabular}
\end{center}
\end{table}

\end{document}